# Non-parametric Local Pseudopotentials with Machine Learning: a Tin Pseudopotential Built Using Gaussian Process Regression

*Johann Lüder,[1] and Sergei Manzhos[2]*

[1] Department of Materials and Optoelectronic Science, National Sun Yat-sen University, 80424, No. 70, Lien-Hai Rd., Kaohsiung, Taiwan, R.O.C. E-mail: Johann.Lueder@gmail.com

[2] Centre Énergie Matériaux Télécommunications, Institut National de la Recherche Scientifique, 1650 boulevard Lionel-Boulet, Varennes QC J3X1S2 Canada. E-mail : sergei.manzhos@emt.inrs.ca

## Abstract

We present novel non-parametric representation math for local pseudopotentials (PP) based on Gaussian Process Regression (GPR). Local pseudopotentials are needed for materials simulations using Orbital-Free Density Functional Theory (OF-DFT) to reduce computational cost and to allow kinetic energy functional (KEF) application only to the valence density. Moreover, local PPs are important for the development of accurate KEFs for OF-DFT as they are only available for a limited number of elements. We optimize local PPs of tin (Sn) using GP regression to reproduce the experimental lattice constants of α- and β-Sn, the energy difference between these two phases as well as their electronic structure and charge density distributions, which are obtained with Kohn-Sham Density Functional Theory employing semi-local PPs. The use of a non-parametric GPR-based PP representation avoids difficulties associated with the use of parametrized functions and has the potential to construct an optimal local PP independent of prior assumptions. The GPR-based Sn local PP results in well-reproduced bulk properties of α- and β-tin, and electronic valence densities similar to those obtained with semi-local PP.

# Introduction

Density Functional Theory (DFT) is a workhorse method of modern computational material science for atomic-scale modeling. Its foundation, the so-called Hohenberg-Kohn theorems,[1] states that the total energy of a system, and therefore other physical observables, can be computed as a functional of the ground state electronic density and is minimized in the presence of an external potential. However, early attempts to compute non-trivial systems such as molecules with a functional of only the density $n(r)$ (such as the Thomas Fermi model)[2,3] were unsuccessful. Introducing orbitals into the evaluation of the kinetic energy, resulting in the Kohn-Sham (KS) formalism,[4] has set the path for the success of DFT. With the Kohn-Sham (KS) ansatz, non-interacting kinetic energy is used:

$$E_{kin} = \sum_i^N \langle \varphi_i | \hat{T} | \varphi_i \rangle \qquad (1)$$

where $\hat{T} = -\frac{1}{2}\Delta$ is the kinetic energy operator and $\varphi_i$ are the orbitals of a system with $N$ electrons. While this formulation has practical advantages such as explicit accessibility of (KS) orbitals and of the band structure and the possibility to include fractions of exact exchange (as used in hybrid functionals),[5] it also sets numerical hurdles for large scale applications with a scaling of approximately $O(N^3)$.

Today's computational materials science with KS-DFT relies to a very significant degree on pseudopotentials or related schemes (such as PAW[6,7]), which can avoid an explicit treatment of core electrons. Since the early successes of pseudopotentials in computational material science, e.g., the formulation by Phillips and Kleinman,[8] many different types of PPs[9] contributed to DFT's success due to reasonable transferability, i.e., they are accurate for many different materials.[10] With the help of developments such as efficient algorithms, basis sets, and PP, the cost can be substantially reduced,[11–16] and order-$N$ scaling in DFT codes can be achieved[11,16–18] but often with large pre-factors[18–20] still limiting routine applications of DFT to systems with few thousand and rarely more atoms, while mesoscopic system sizes are beyond the capabilities of KS-DFT. This leaves outside the scope of KS-DFT numerous phenomena which are intrinsically large-scale such as microstructure-driven properties or dynamics of biological systems (protein folding, transport in ion channels, etc.).

In contrast to KS-DFT, Orbital-free Density Functional Theory (OF-DFT) computes the total energy as a functional of the total electronic density. Then, the total and kinetic energy do not explicitly depend on individual electronic orbitals but are expressed as functionals of density-dependent variables (values of the density or any derivatives and powers thereof). The total energy can be written in functional form as

$$E_{total} = E_{kin}[n] + E_H[n] + E_{XC}[n] + E_{ext} \qquad (2)$$

where energy components (Hartree energy $E_H$, exchange-correlation energy $E_{XC}$, and the energy in the external potential $E_{ext}$) can be the same as in KS-DFT, and $E_{kin}[n]$ is the kinetic energy functional (KEF). This brings OF-DFT closer to the Hohenberg-Kohn theorems than KS-DFT and it can lead to a significant reduction of computational cost. Near-linear scaling can be obtained with small prefactors.[21,22] Because of the scaling advantages, OF-DFT is a promising way to circumvent the computational limitations of KS-DFT; with it, simulations of physical and chemical properties of large atomic structures reaching the mesoscopic scale can be achieved, and calculations with a few dozen thousands of atoms are feasible on a consumer desktop computer.

Note that with Eq. (2) $E_{kin}[n]$ should be equal to the non-interacting KS kinetic energy, i.e. Eq. (1), with correlation contributions to the kinetic energy handled via the $E_{XC}$ term. Modern KEFs[23–29] work well for systems which have relatively homogeneous densities. Larger fluctuations of the density than in those materials, such as in transition metals with partly occupied *d*-orbitals, electronic shell structure in an atom or with directional / covalent bonds, lead to failure of these functionals. Thus, the development of a universal kinetic energy functional, or even of functionals which would work well for different types of systems with more complicated valence structures, is still one of the most important obstacles for OF-DFT.[24,30–34]

For large-scale systems, valence properties are of interest in most cases, as core-level properties and core-level excitations are mostly of local nature and, if they are of interest, can usually be computed with small-scale models within KS-DFT. With this in mind, the core-levels can be ignored by replacing atoms with pseudoatoms, thereby avoiding the complicated inner shell structure with large density fluctuations in OF-DFT calculations. These fluctuations are still a challenge for reliable calculations with kinetic energy functionals, as existing approximations work best on slowly varying densities.

Using pseudoatoms requires working with pseudopotentials. With good pseudopotentials, practically no sacrifices in accuracy of valence properties occur. On the other hand, most modern pseudopotentials were developed for KS-DFT and are non-local (or semi-local) pseudopotentials which are dependent on orbital angular momenta that cannot be used in OF-DFT due to the absence of orbitals. OF-DFT, therefore, requires accurate local pseudopotentials. Local pseudopotentials (LPP) typically reach only the accuracy of non-local (NLPP) and semi-local pseudopotentials (SLPP) for materials with a relatively homogenous electronic density such as Li, Na or Ag.[35–37] This limitation results in another prominent obstacle for progress in OF-DFT. Accurate local PPs are not available for most elements of the periodic table. Only a handful of local PPs for light metals and only a couple covalent systems exists today.[35,36,38–42] This not only limits the use of existing OF-DFT approximations but also inhibits the development of new KEFs, as those are more likely to succeed if they operate on the smoother valence density. This is therefore a major and relatively underappreciated issue in OF-DFT development.

Improving both local PPs and KEFs faces many challenges within the framework of OF-DFT. For instance, there are compounding errors because the quality of one - the local PP or the KEF - will affect the development and the quality of the other one. Local pseudopotentials are generally less accurate than non-local pseudopotentials. Here, we are concerned with constructing the best possible LPP within the limits imposed by the LPP approximation.

One approach to this problem is that the construction of accurate local PPs that yield accurate valence properties, including atomic structures and density, should be sought outside OF-DFT. Recent developments of local PPs employ KS-DFT for this purpose. Carter and coworkers developed a method to obtain a local PP through inversion of electronic density computed with KS-DFT which produced accurate pseudopotentials for, e.g., Li, Mg and Al.[36,39] This has, however, the disadvantage that the PP is derived using one DFT setup (KS-DFT with KS KEF, a specific basis set, etc.), while the local PP is used with another computational framework (OF-DFT with a specific KEF, density expansion, etc.). This results in compounding errors of KS- and OF-DFT. An alternative approach is to fit a local PP within a given OF-DFT setup to avoid this compounding of errors. For example, Legrain and Manzhos optimized the parameters of empirical functions to represent local pseudopotentials for Li, Na and Mg to fit observables such as structures or relative phase energies with OF-DFT.[35]

While these two approaches have produced excellent local PPs for a few elements (e.g. Li, Na, Mg and Al), further developments of local PP based on simple parameterized functions are difficult. For instance, imposing a parameterized functional form may prevent achieving the best local PP, and the need for an initial guess (of the functional form and parameters) may exclude the best solution of an optimization scheme. We would like a robust approach to build PPs that is flexible enough to accommodate any PP shape, to circumvent any form of error propagation, and is independent of an initial guess/assumption. This approach could also include the use of reliable experimental and theoretical data (such as structures, electron densities, etc.) to construct local PP.

In this work, we use machine learning of a parameter-free PP representation to achieve this goal. Machine learning has become a promising tool in computational material science for materials discovery and pre-screening for, e.g., pharmaceuticals,[43,44] representing potential energy surfaces[45,46] as well as in electronic structure.[47–54] Here, we employ Gaussian Process Regression (GPR)[55,56], which is a supervised statistical learning method that optimizes an unknown function over a training set. GPR can be applied to a broad range of regression problems. It is a parameter-free approach (only hyper-parameters are used) and, therefore, a flexible representation that can accommodate any local PP shape. In addition, GPR does not restrict the search space in any optimization scheme. This also implies that no prior guess of a parameterized functional form is required (although a starting point for the PP optimization is required). In GPR, only the kernel, which is a covariance function, is evaluated with hyper-parameters, which can be optimized as well. As a result, the GPR representation of a local PP is not tied to any particular method to generate data such as a specific DFT setup or measurements but it provides the possibility to optimize the local PP to yield accurate results with DFT for any material. *GPR can, in principle, be used to construct the best possible local PP.*

This approach is used to construct a local PP of tin (Sn). Sn is an interesting material because it has many commercial applications and is, at the same time, a challenge for *ab initio* calculations due to its phase transition at 286 K at low pressure from α- to β-Sn (corresponding to a difference in energy per atom of only about 20 meV between β- and α-Sn). This small energy difference challenges computational methods (many DFT implementations compute a much larger phase energy difference) and can be a target for development efforts of more accurate theoretical methods and a better understanding of those.[57] This is important when modeling phenomena involving phase transitions, such as lithiation of Sn for Li-ion batteries.[58,59] Simulations of such phenomena of practical importance would also significantly benefit from large-scale simulations in particular with OF-DFT. The α and β-phases of Sn are shown in Figure 1.

Further applications of Sn can be found in semiconductors such as in SiSn, GeSn and $SnS_2$ alloys where it can be used for bandgap engineering of materials, in piezoelectric and infra-red devices.[60–62] It is also used in commercially available superconducting magnets (e.g., $Nb_3Sn$),[63,64] nuclear fuels (e.g. $Zr_ySn_x$)[65,66] and in optoelectronic devices[67–69] as well as nanostructured materials,[70,71] in addition to the earlier mentioned applications in Li-ion batteries.[58,59] Providing a local PP of Sn for accurate computations of the physical and chemical properties of Sn and Sn alloys, including the Sn phase separations, could fundamentally accelerate further technological advances based on microscopic and mesoscopic Sn-systems.

In this study, we thus present new representation math for local PP based on GPR. We optimize the local PP for Sn empirically to reproduce reference data from experiments and DFT computations with NLPPs such as lattice constant, heat of formation, electronic valence density, eigenstates and relative cohesive energy between different phases of a material.

## Methods

**Gaussian process regression**

For the GPR representation of a local PP as a function $y(x)$, we adopt GPR in the general form[46,55,56]

$$\bar{y} \sim N(\bar{0}, \underline{k}(\bar{x}, \bar{x})), \qquad (3)$$

which expresses a joint multivalent Gaussian distribution $N$, where $\bar{x}$ is the vector of radial points of the local pseudopotential, $\bar{y}$ is the vector of the potential values at positions corresponding to $\bar{x}$, $\underline{k}$ is a covariance matrix measuring similarity between points in $\bar{x}$. The distribution is assumed to have a mean of 0. $(\bar{x}, \bar{y})$ constitute a training set. When the covariance matrix is determined on a training set of the data, for value $y'$ of any point $(x', y')$ not included in the training set can be estimated as

$$\begin{bmatrix} \bar{y} \\ y' \end{bmatrix} \sim N \left( \bar{0}, \begin{matrix} \underline{k}(\bar{x}, \bar{x}) & \bar{k}'^T(x', \bar{x}) \\ \bar{k}'(\bar{x}, x') & k(x', x') \end{matrix} \right) \tag{4}$$

where $\bar{k}$ is a covariance vector of $\bar{x}$ and $x'$ and $k$ is simplify evaluating $k$ locally on $x'$. An error interval (variance) can be estimated for $(x', y')$ according to

$$\Delta y' = k - \bar{k}' \underline{k}^{-1} \bar{k}'^T \tag{5}$$

since a Gaussian distribution of $y'$ is assumed. Given $(\bar{x}, \bar{y})$, GPR essentially produces Bayesian estimates of values at any $x$ (which are later adapted as radial distances from the core of a pseudoatom). Therefore, it allows computing the value of the pseudopotential at any point in space without having a pre-determined functional form and avoids restrictions associated with it. Although Eq. (4) implies a continuous function for $y'$ (a weighted sum over functions $k$), its form is not pre-determined. Literature suggests that GP regression is as good or better than other universal approximators such as neural networks.[45,72]

**Fitting of local pseudopotentials with Gaussian processes regression**

The optimization of the local Sn PP is performed in four steps as described below and the generation of the potential (steps one and two) is illustrated in Figure 2. The initial guess of the local PP of Sn (Z=50) is obtained via interpolating previously published local PP of In (Z=49) and Sb (Z=51),[73,74] see Figure 2 (a).

We use GPR implementation in the Scikit-learn package[75] for Python. GPR is used to represent the non-asymptotic region near the core:

$$V(r) = V_{GP}(r)[1 - \sigma(r - R_m)] + \frac{Z_v}{r} \sigma(r - R_m) \tag{6}$$

where $Z_v$ is the valence charge, $V_{GP}(r)$ is the GPR-represented function, $\sigma(r)$ is a sigmoid switching function, $\sigma(r) = (1 + \tanh(ar))/2$ with $a = 1$ Bohr$^{-1}$, and $R_m$ is the center of the switching function between the core and the asymptotic regions. The switching function is shown in the inset of Figure 2(a). This ensures a smooth transition between the regions.

The local PP is encoded by a set of values $\{V_{GP}(r_i)\}$ with $i = 1, ...., n$ on a (non-equidistant) set of points $r_i$. These values form a parameter string. First, the potential values are rescaled so that the maximum and the minimum take values of +1 and -1, respectively. Then, a minimal set of training points $\{r_i\}$ is searched to represent the local PP with GPR. With an initial choice of $n = 12$ points, a random selection from $r$ values on a fine grid ($r$ =[0,16] Bohr with a resolution of 0.01 Bohr, giving a total $N$=1600 values) is made and the training values for the GP regression are read from the initial local PP at the $n$ points. The fit of the potential values is obtained according to Eq. (3-5) with the radial basis function kernel

$$k(r_i, r_j) = exp(-\frac{\|r_i - r_j\|^2}{\gamma^2}) \qquad (7)$$

that measures the correlation between two points. Given the correlation function and the training set, the fitted potential is computed for all other points in $r$, .e.g., on the entire grid of $N$ points with GPR. The length scale $\gamma$ of the kernel is optimized in a range from 0.3 to 2.0 to minimize the root-mean-square error (RMSE) which is evaluated as residual difference between the initial potential values ($V$) and its fit ($f$) over all $N$ points in $r$ as

$$RMSE = \sqrt{1/N \sum_i (V(r) - f(r))^2}. \qquad (8)$$

Preliminary tests to estimate the influence of small changes (errors) in the local PP on properties computed with DFT (see below) are performed. For this purpose, the interpolated local PP was modified with small contributions from a fit near the pseudo nucleus. Since a GP regression gives values $V$ for each training point $r$ and a value $V$ and error estimate $\Delta V'$ for all other points ($r'$), a scaled $\Delta V'$ was added to the initial local PP at all $r'$ in an interval between two neighbored $V$'s near the potential minimum. This leads to a controlled modulation of RMSE with respect to the initial local PP. We find that a RMSE of less than 1/5,000 Ha (later referred to as RMSE threshold) leads to negligible changes in the charge density and in changes in bulk properties. Examples of fits with different $\Delta V'$ are shown for several RMSE in Figure 2 (b) and its inset. Similar values of RMSE relative to the energy range are known to yield accurate GP fits for other types of potentials such as potential energy surfaces.[45] For the actual GPR fit of the local PP, the random selection of training points is repeated 20 times and if the RMSE threshold is not reached then, the number of training points is increased. This is repeated until the RMSE threshold is reached which is generally achieved with 30 to 60 training points in our fitting routine for the local PP of Sn. After successful completion, this yields the GPR fit of the local PP and the training set, which is used in the next step.

**Potential variation**

In step two, an optimized local PP is searched through small changes in the data points of the determined training set with random-walk type variations of the training values $\bar{V}$ that are within a core region radius $R_m$ from the pseudo core. Outside $R_m$, the PP should follow the asymptotic behavior of the Coulomb potential while inside $R_m$ the potential should mimic the interaction with core electrons (those that are not treated explicitly) and the unscreened nucleus on the valence electrons. For each $r$ within $R_m$, the corresponding $V$ is randomly (uniform distribution) changed by values between -0.05 and 0.05 Ha and a new modulated local PP is generated with the new set of data points that contains the modulated data pair via the same GPR routine as described above using the same $\gamma$ as the initial fit. The new local PP is tested to confirm that it has only small changes from the initial fit $f$. The RMSE between $f$ and the modulated local PP must be less than 0.4 Ha and all points of the new local PP are within a band spanning a region of $\pm$ 0.2 Ha around $f$ at each point $r$. If these conditions are not fulfilled, a new fit is started. Otherwise, the new potentials are rescaled from the range [-1,1] back to the original scale. Figure 2 (c) shows a fit of the initial potential with more than 60 data points resulting in a fit with a RMSE less than 1/5,000 Ha and the new potential that is modified at its minimum. This step results in several modulated local PPs that is equal to the number of training points within $R_m$.

**Density Functional Theory computations**

In step three, DFT calculations on α-, β-Sn and a single atom Sn are performed for each generated local PP. The DFT calculations are plane-wave based and performed with the *Abinit* code.[76–80] The Perdew-Burke-Ernzerhof (PBE) exchange-correlation functional[81] is used. The self-consistent field cycle was stopped when the energy reached convergence taken as energy difference of less than $3\times10^{-8}$ eV is reached in two consecutive iterations. Single point calculations for single atom systems, i.e., fixed geometry, as well as volume and ionic relaxations for bulk systems, were performed. When structure optimizations were allowed, volume relaxations were performed until all forces on the atoms were less than $1.0\times10^{-3}$ eV/Å and isotropic pressure was below $10^{-4}$ GPa.

Fermi-Dirac smearing was used as broadening function with σ = 50 meV as broadening parameter if not stated otherwise. Because we are aiming to reproduce bulk material properties using local PPs with high accuracy to reference data, e.g., valence density distributions, convergence criteria are of critical importance. The k-point sampling the Brillouin-zone of the bulk materials used the Monkhorst-Pack scheme and the sampling uses a k-point density corresponding to at least $k \times a_0 = 33$ Å ($k$ being the number of k-points and $a_0$ the length of a lattice vector). For the single atom calculations, only the Γ-point and a box size of 12×12×12 Å are used. The plane wave cutoff was 1,600 eV. This computational setup yields well-converged physical quantities, including density contributions with non/semi-local PPs.

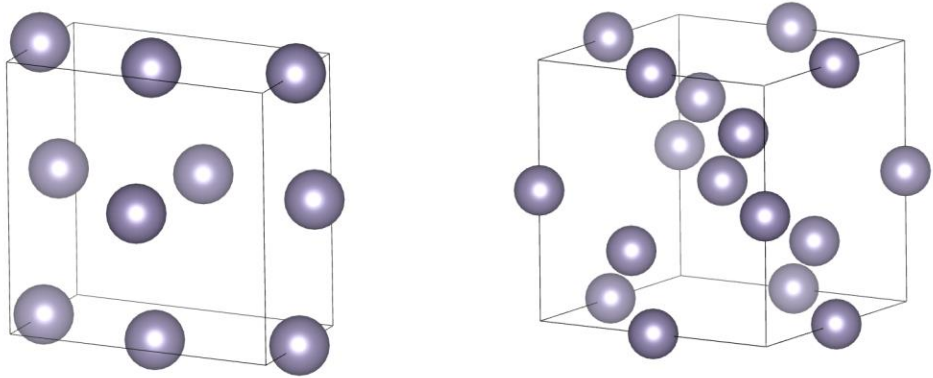

Figure 1. Unit cells of β-Sn (left, Materials Project[85] structure ID mp-88) and α-Sn (right, Materials Project structure ID mp-117).

The initial unit cell parameters of α- and β-Sn are taken from experimental references.[82–84] The initial structure of α-Sn is a cubic unit cell with a lattice constant of 6.48 Å and the initial β-Sn structure (tetragonal lattice) with lattice constants $a_0 = 5.82$ and $c_0 = 3.18$ Å. These values also serve as reference values for the structures of the optimized unit cells in the GPR-optimization scheme. The structures are shown in Figure 1.

Several SLPPs are used to compute reference properties of α and β-Sn. We use *Abinit's* Fritz-Haber Institute Troullier-Martins[86] (FHI), the Goedecker-Teter-Hutter[87] (GTH), and the optimized norm-conserving Vanderbilt[80] (ONV) SLPPs. These potentials are semi-local in the orbital channel (i.e. momentum space) while they are local in real space. The FHI and the GTH PPs have the same valence-electron configuration for Sn, i.e. four electrons in the 5$^{th}$ electronic shell, as our *to-be-optimized* local PP of Sn. The ONV PP is another SLPP which accounts for 14 electrons including valence and semi-core electrons in a Sn atom. The additional semi-core electrons should result in a more accurate description of physical and chemical properties based on the electronic states, e.g. mildly affecting the lattice constants and the cohesive energies used as reference data. However, the (pseudo) valence density obtained with the ONV is not directly comparable to the one computed with our local-PP or the other two SLPPs. The chosen SLPPs provide an estimate of accuracy for physical properties computed with the GPR-optimized local PP, e.g., the valence-only density distribution of single atoms or bulk materials additionally to values that can be compared to experimental values such as lattice constant and put that accuracy in perspective.

**Evaluation of the accuracy of the local pseudopotential**

In step four, the computed DFT properties of α-, β- and the single atom Sn are evaluated for each of the newly generated local PPs. The relative errors of computed lattice constants of the bulk structure are evaluated according to

$$A = \sum_i \|a_i - \tilde{a}_i\| / \tilde{a}_i \qquad (8)$$

where $a_i$ is the computed lattice constant of direction $i$ and $\tilde{a}_i$ is an experimental/theoretical reference value. The electron density of the unit cells and the single atom calculation are computed on a grid of $128^3$ points. The density difference ($\Delta\rho$) between the same structures computed with different PP is evaluated according to

$$\Delta\rho = \sum_{i,j,l}^{All} \|\rho_{i,j,l} - \tilde{\rho}_{i,j,l}\| / \sum_{i,j,l}^{All} \tilde{\rho}_{i,j,l} \qquad (9)$$

where $\rho$ and $\tilde{\rho}$ are the densities obtained with a local and a SLPP, respectively. Their difference is taken at all grid points defined by their principal components $i, j$ and $l$, which are evaluated over the three lattice directions over all points of the $128^3$ grid, respectively. The difference in the eigenvalue spectrum ($\Delta\xi$) is evaluated relative to the valence band maximum (VBM). Then, the energy separation between the computed states over momentum space is

$$\Delta\xi = \sum_{i,k}^{N,K} \|\varepsilon_{i,k} - \tilde{\varepsilon}_{i,k}\| / \sum_{i,k}^{N,K} \|\tilde{\varepsilon}_{i,k}\| \qquad (10)$$

where $\varepsilon_{i,k}$ and $\tilde{\varepsilon}_{i,k}$ are computed and reference values, respectively, of the $i^{th}$ eigenvalue at the $k^{th}$ k-point of the Brillouin zone of the occupied spectrum. For the eigenvalues of single atom calculations, the summation over k-points is dropped in Eq. (10). Additionally, the summations include the conduction band minimum (CBM) to also take into account the bandgap (in α-Sn). The absolute cohesive energy per atom $E_a^c$ (that is an approximation of the heat of formation neglecting entropy and vibrations) of a material is computed as the total energy difference between the unit cell of the material and all its single atom components according to

$$E_a^c = \frac{E_M}{n} - E_s \qquad (11)$$

where $E_M$ is the total energy of the simulation cell with $n$ atoms and $E_S$ is the energy of an isolated Sn atom. The relative cohesive energy (or formation energy) $E_r^c$, e.g., between phases $p_1$ and $p_2$ (e.g. between α- and β-Sn), is computed as total energy difference per atom of these structures according to

$$E_r^c = \zeta_{p2} - \zeta_{p1} \tag{12}$$

where $\zeta_{p1}$ and $\zeta_{p2}$ are the total energies per atom for phase $p_1$ and phase $p_2$, respectively. The difference of relative cohesive energy ($\Psi$) due to use of different PPs is computed as

$$\Psi = E_r^c - \tilde{E}_r^c \tag{14}$$

where $\tilde{E}_r^c$ is a reference value (such as with the semi-local FHI PP or an experimental value).

Finally, the local PP of Sn is updated with one of its modulations if the penalty function P defined as

$$P = \sqrt{\omega_1 \Psi^2 + \omega_2 \left(E_a^c - E_a^{c,exp}\right)^2 + + \sum_j \left(\omega_{j,3} \Delta \xi_j^2 + \omega_{j,4} A_j^2 + \omega_{j,5} \Delta \rho_j^2\right)} \tag{15}$$

is lowered. Here, the summation over j includes all phases of Sn and the single atom configuration while $\omega_i$ are weights (which are deemed to have units of the inverse of respective terms). The weights were chosen to make all terms in the sum have similar order of magnitude. This is so that their relative errors are comparable. For instance, a density difference of 2% should result in a summand of comparable magnitude to that due to a lattice mismatch of 2%, or to an error in absolute cohesive energy that, based on the DFT results, can be in the order of 100 meV. This form takes into account the density residuals in the bulk materials and in an isolated atom, the relative and absolute cohesive energies, the residual in the eigenvalue spectrum and the difference in lattice vectors.

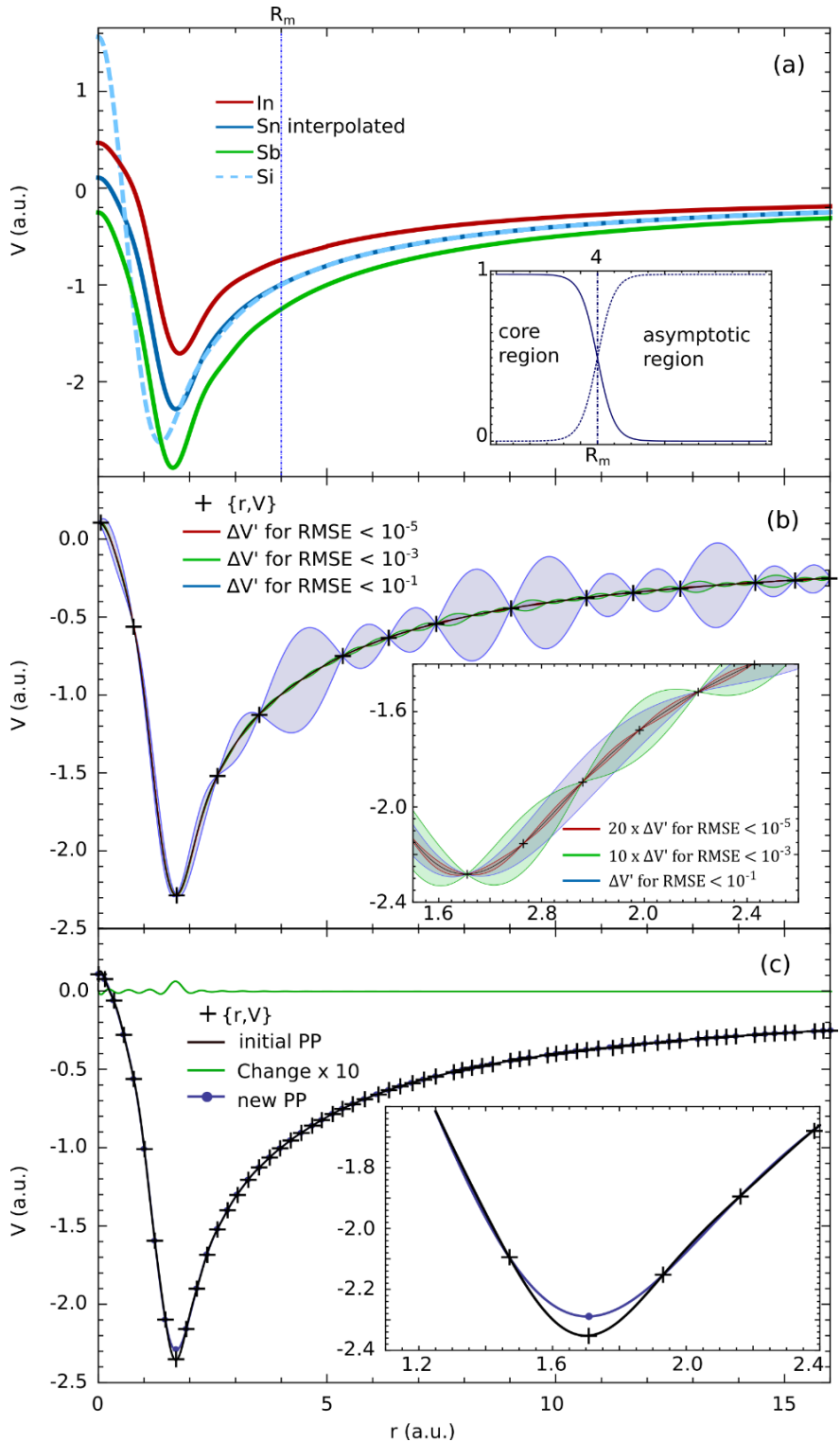

Figure 2. (a) Interpolation of the local Sn PP from the local PP of In and Sb73,88 and comparison to the local PP of Si.36 The inset shows a possible switching function between core and valence, i.e., asymptotic region. (b) Standard deviation ($\Delta V'$) and GPR fit of the local Sn PP that is scaled to potential values between 1 and -1 to perform the GP fit and then rescaled for plotting. The training points {r,V} are given of the fit with RMSE < $10^{-1}$ while the inset shows the training points {r,V} for RMSE < $10^{-5}$. (c) Plot of a GPR fit of the local Sn PP potential with RMSE < $10^{-5}$ (blue curve) and one of the modified PP (black curve) for which the change (green curve) was multiplied by 10. The inset shows a magnification of the modified training point and the resulting changes in the local PP. Rm is at 4 Bohr. RMSE reported in Ha.

## Results and discussion

**Interpolated local PP of Sn**

The interpolated local PP of Sn is shown in Figure 2 (a), together with the local PP of Sb and In (that are used for the interpolation) in addition to the PP of Si, for comparison. The PP of Si appears to be similar to the one of Sn because the PP of these elements have a similar valence configuration between. Beyond $r$ = 4 Bohr, the local PPs of Si and of Sn follow a similar asymptotic behavior while displaying differences in the core region due to different core sizes (taken as $R_m$). In the core region, the minimum of Si's PP is by about 0.4 Bohr closer to the core (and slightly deeper) and with a larger maximum at the core resulting in a steeper potential slope. Similarities between the local PP of Sn and the local PPs of In and Sb, that are a minimum at ca. 1.6 Bohr, a potential value close to 0 at the core and a similar asymptotic behaviour at large r, are imparted to the interpolation.

Consequently, charge densities obtained with the local Sn PP experience weaker potential-based repulsion from the core at distances between 0 to ca 1 Bohr, and a reduced attraction between 1 and 2 Bohr. At distances larger than 2 Bohr, the local potentials of Sn and Si are very similar to each other. This agrees with a somewhat larger spatial extension of the valence wavefunctions of Sn compared to the ones of Si as well as the larger atomic radii ($r_a$(Si) = 2.32 Å and $r_a$(Sn) = 2.48 Å).[89] Therefore, the initial guess of a local PP of Sn taken as interpolation of the local PP of Sb and In is, at least qualitatively, reasonable and a valid starting point for our empirical optimization scheme. While the qualitative picture drawn by the potential appears promising, quantitative results, such as phase ordering and cohesive energy, obtained with the interpolated local PP of Sn demonstrate the need for substantial improvements, as shown below.

**GPR PP of Sn**

The final GPR-optimized PP of Sn is shown in Figure 3. Two PPs are shown, labeled LPPv1 and LPPv2. Because a well-defined criterion to select a radius that splits the core from the valence region for local pseudopotentials is missing, we test two different values for $R_m$ and compare the effect on the GPR-optimized local PP. We choose 4 Bohr for one of the local PP (LPPv1), while 8 Bohr allows a more flexible optimization of the other local PP (LPPv2). Besides this difference, the same optimization process, including taking the initial guess as interpolated PP, was used.

Both GPR-optimized PPs are very similar to each other. The most noticeable differences are in three regions, as highlighted by the insets of Figure 3: 1.5 to 1.9 Bohr, 3 to 4 Bohr and 6.0 to 6.6 Bohr. The largest difference is in the middle region (3 to 4 Bohr). This region marks the center of the switching function for LPPv1. Hence, this feature is suppressed for LPPv1, which also applies to the outer region (6.0 to 6.6 Bohr), while the difference seen in the inner part might partly compensate for other effects. It should be highlighted that the near-core region (ca. $r < 1$ Bohr) is barely affected.

**Structure, energies and phase ordering of α and β tin**

In Table 1, we compare the computed lattice constants of α- and β-Sn, and the relative and absolute cohesive energies for the semi-local PPs, i.e. FHI, GTH and ONV, to the results obtained with the interpolated local PP, and the GPR-optimized PPs, as well as to experimental values. In addition, values computed with the VASP package[90,91] are taken from the Materials Project.[85]

The SLPPs overestimate the lattice constant of α-Sn by ca. 3 %. This is an expected result due to the employed PBE exchange-correlation functional, as it is confirmed by all-electron calculations employing a full-potential linearized augmented plane-wave and local orbitals method used in Ref. [92] by Haas *et al.*, where a lattice constant of α-Sn is reported to be 6.66 Å, is 2.6 % larger than the experimental value. Surprisingly, the interpolated local PP yields a much better agreement of the lattice constant for both α- and β-Sn, with agreement better than 99 %, than some of the SLPPs. Also, the GPR-optimized LPPs yield very good (LPPv2) and acceptable (LPPv1) lattice constants for Sn. The max. rel. errors are 3.26 % for LPPv1 and 2.4 % for LPPv2. Still, the well-reproduced experimental lattice constants obtained by the local PPs might be misleading and, to some amount, artificially counteracting intrinsic limitations of the employed exchange-correlation functional.

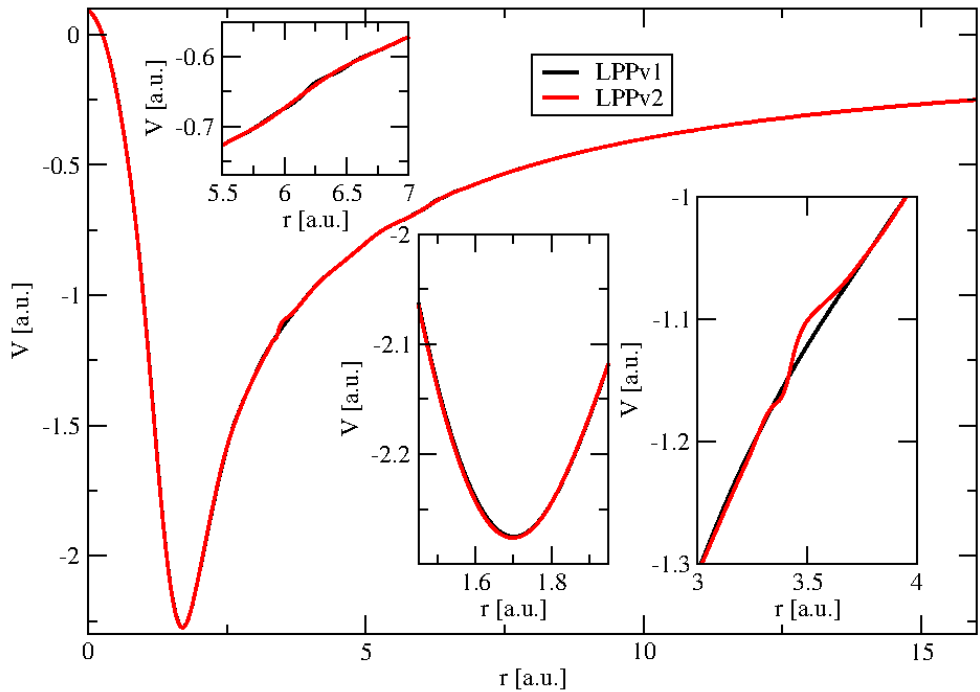

Figure 3. The shape of two GPR optimized local pseudopotential of Sn with different $R_m$.

Table 1. Comparison of experimental and computed reference values for α- and β-Sn lattice parameters $a_0$ and $c_0$, the cohesive energy of α-Sn $E_a^c$ and relative cohesive energy of β- vs α-Sn $E_r^c$ to computed values obtained with different SLPPs (FHI, GTH and ONV), the interpolated local PP, and the GPR-optimized LPPs. The most accurate values obtained with the local PP are highlighted in bold font.

| | | $a$ (Å) | $c$ (Å) | $E_a^c$ (eV) | $E_r^c$ (meV) |
|---|---|---|---|---|---|
| Exp./Ref. values | α-Sn | 6.46 – 6.4912 | -- | -3.14 | 20 |
| | β-Sn | 5.8197 - 5.8316 | 3.1749 - 3.1815 | | |
| FHI | α-Sn | 6.64 | -- | -3.5 | 13 |
| | β-Sn | 5.97 | 3.14 | | |
| GTH* | α-Sn | 6.67 | -- | -3.43 | 39 |
| | β-Sn | 5.90 | 3.22 | | |
| ONV* | α-Sn | 6.66 | -- | -3.46 | 40 |
| | β-Sn | 5.93 | 3.23 | | |
| PAW[a] | α-Sn | 6.65 | -- | -- | 47 |
| | β-Sn | 5.93 | 3.22 | | |
| FP-(L)APW+lo[92] | α-Sn | 6.66 | -- | -- | -- |
| Interpolated local PP[b,*] | α-Sn | 6.50 (0.2%) | -- | -3.51 (11.7%) | 134 |
| | β-Sn | **5.86 (0.5%)** | **3.19 (0.3%)** | | |
| GPR PP (LPPv1)* | α-Sn | 6.51 (0.3%) | -- | **-3.36** | -3 |
| | β-Sn | 6.02 (3.3%) | 3.28 (3.1%) | **(6.5%)** | |
| GRP PP (LPPv2)* | α-Sn | **6.47 (-0.3%)** | -- | -3.45 | **43** |
| | β-Sn | 5.97 (2.4%) | 3.26 (2.5%) | (9.9 %) | |

[a] Data taken from Ref [57].

[b] Taken as an interpolation from the local pseudopotential of In and Sb.[73,88]

*Computed with k × $a_0$ of ca. 80 Å for bulk systems and with σ=5 meV.

All semi-local and local PP here yield $E_a^c$ values between -3.5 and -3.4 eV, which are overestimating the experimental binding strength (-3.14 eV) by ca. 0.4 eV. The relative cohesive energies $E_r^c$ computed with the SLPP have values between 13 to 50 meV. These values agree relatively well to the experimental value of ca. 20 meV. The interpolated local PP, however, computes a value of 134 meV, which is unacceptably overestimated by more than 100 meV. Nonetheless, it must be acknowledged that the relative order of α- and β-Sn phases is computed correctly; that is, α-Sn is more stable than β-Sn (below 286 K). In contrast, the GPR-optimized local PP of Sn obtained relative cohesive energies being much closer to the reference: -2 meV for LPPv1 and 41 meV for LPPv2. While the former result is qualitatively wrong, i.e., β-Sn is more stable than α-Sn at low temperatures, the latter result is of similar accuracy as the ones computed with SLPPs. In fact, LPPv2 obtained the best agreement among the tested local PP for phase ordering while maintaining excellent agreement in the lattice constants. A value of about 40 meV for $E_r^c$ was also reported with the PBE functional when using different (KS) DFT codes and basis types.[57] Considering the technical and physical restrictions given by the local nature of the potential, this result is encouraging. In addition, LPPv2 yields an absolute cohesive energy that is still comparable to the ones obtained by other pseudopotentials and the PAW method. These points indicate that LPPv2 might be very suitable for simulations of structural properties and processes than the other local PPs. It should be noted that these results are obtained for denser k-point sampling (k × a0 of ca. 80 Å) since the convergence of the total energy of less than 1 meV for β-Sn requires 17×17×17 k-points for the primitive unit cell, while structure parameters converge with significantly sparser k-point densities.

**Band structures of α and β tin**

The densities of states (DOS) obtained with a GPR-optimized local PP (LPPv1) employing the tetrahedron integration method of the Brillouin zone[93] are shown in Figure 4 for α- and β-Sn. Band structures are also shown. For comparison, the electronic structure obtained with the semi-local ONV and GTH PP are given. α-Sn has was characterisized as zero-bandgap semiconductor[94], while other earlier experiments results also reported for α-Sn is 0.08 eV[95]. Besides this difference, it is clear that an accurate PP should produce a similar value (a zero or at least a very small bandgap) with a reasonable electronic broadening.

The SLPPs (GTH and ONV) yield almost identical DOSs and band structures for both α- and β-Sn. For these two PPs, α-Sn is computed to have no bandgap. For LPPv1, the band structure is qualitatively correct for both phases, and main features in the DOS and in the band diagram are well reproduced compared to the results obtained with the semi-local pseudopotentials. By comparison, the local PP based electronic structures show a contraction of states that increases with distance to the Fermi energy. Interestingly, a bandgap opens in α-Sn when computed by LPPv1 and LPPv2. LPPv1 computes a direct bandgap of 0.13 eV for α-Sn, which is in satisfactory agreement with the experimental value of 0.08 eV reported by Ewald *et al.*. However, LPPv2 yields an increased energy gap at the Γ-point, see Figure *5*. Besides, α-Sn changes with LPPv2 to an indirect bandgap material with a valence band maximum at Γ and a conduction band minimum at *L*. The indirect bandgap is 0.68 eV and direct bandgap is 0.93 eV. This is only slightly different from the values obtained with the interpolated potential, i.e. 0.63 and 0.82 eV for indirect and direct bandgap, respectively. Also, the DOS and band structure of β-Sn is contracted for the local PPs when compared to SLPPs. The difference in band structure for β-Sn, which is correctly metallic for both optimized versions of the local PP (and the SLPPs), is small when computed with LPPv1 and LPPv2. These results indicate that electronic properties are better reproduced with LPPv1 between these local PPs.

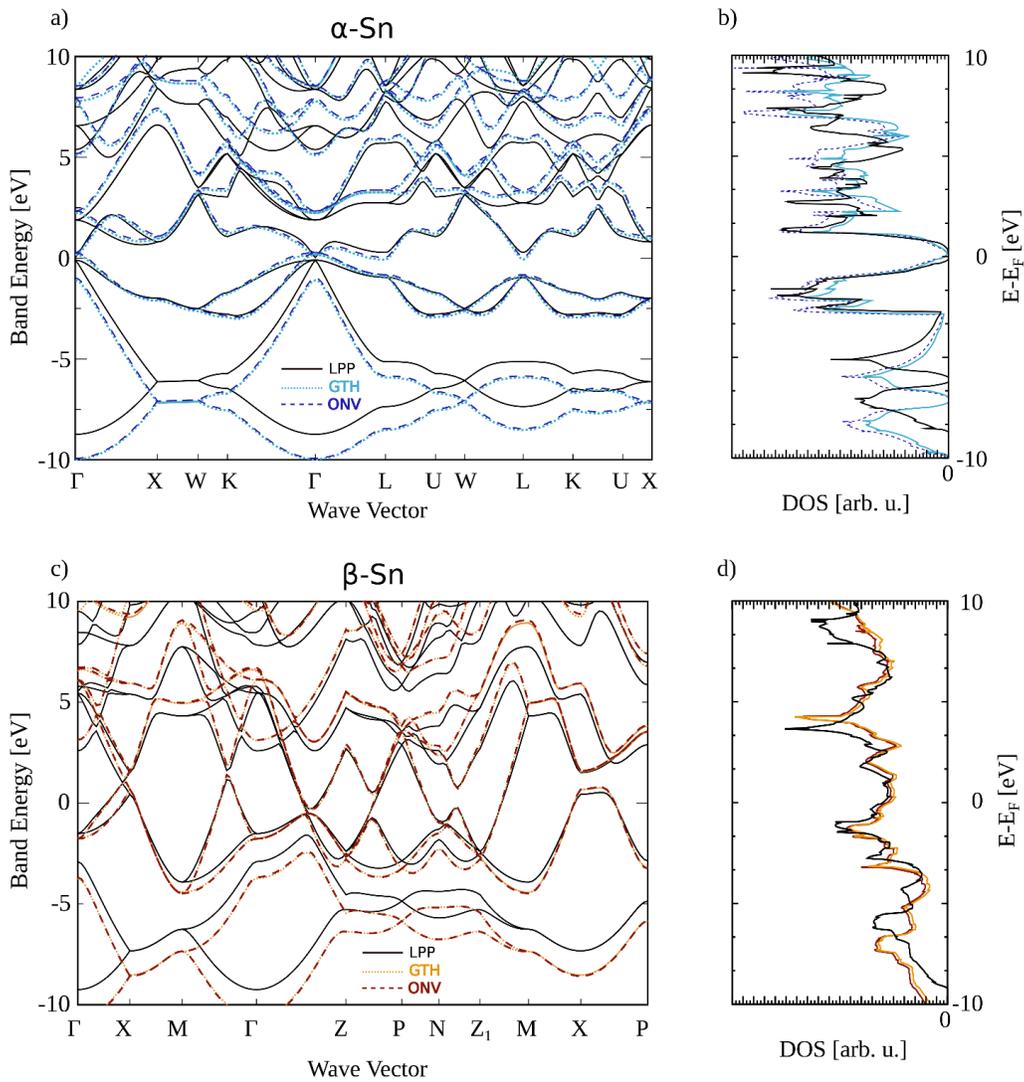

Figure 4. a) The band structure and b) the density of states of α-Sn obtained with the GPR fitted local pseudopotentials (LPPv1) in comparison to those obtained with the GTH and ONV pseudopotentials. c) The band structure and d) the density of states of β-Sn obtained with the GPR fitted local pseudopotentials (LPPv1) in comparison to those obtained with the GTH and ONV pseudopotentials.

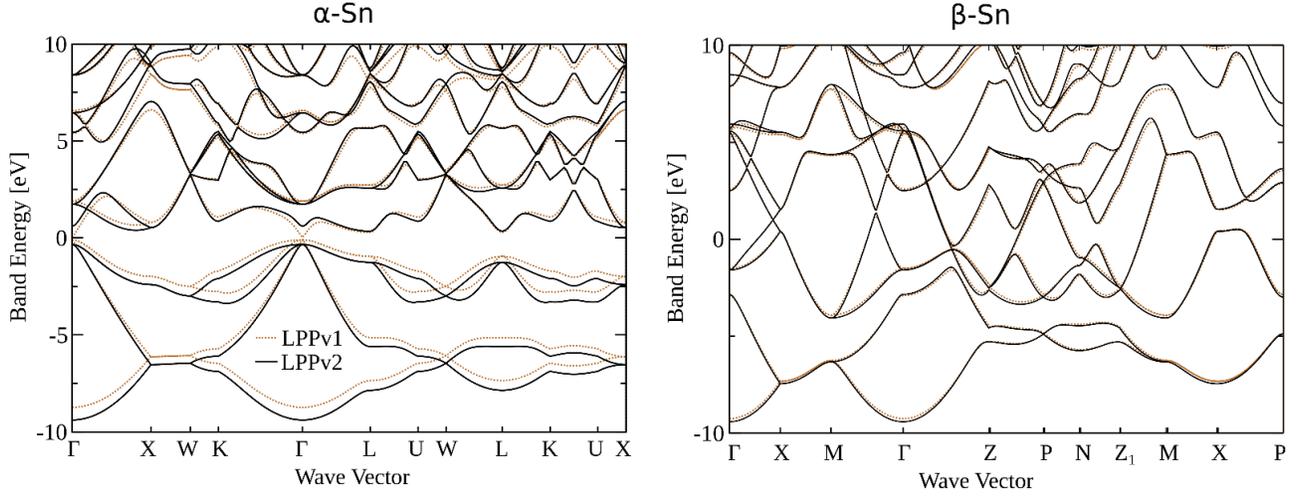

Figure 5. Comparison of band structures of α- and β-Sn obtained with LPPv1 and LPPv2. The band structures are aligned at the Fermi energy.

**Electron densities**

Table 2 compares the relative density difference for a selected phase computed as $\delta\rho = \int(\tilde{n}_a(r) - \tilde{n}_b(r))dV$, where $\tilde{n}_i$ are normalized densities computed with two different potentials, indicated by the indices $a$ and $b$, that can be semi-local or local PP with the same number of valence electrons. As a reference density cannot be reliably defined among all the pseudopotentials considered, we compute differences in density between all pairs of pseudopotentials.

For α-Sn, the SLPPs show little differences in density (less than 2 %). As expected, the interpolated PP has a larger difference compared to the SLPPs' results. This is partly compensated during the optimization of LPPv1, which reduces the difference from about 5% to about 3%. In contrast, the larger $R_m$ in LPPv2 and the resulting modification in potential shape increases density residuals (4%). The results follow, in general, similar trends in densities of β-Sn; however, LPPv2 results in densities closer to those of SLPP than LPPv1.

Interestingly, the FHI PP and the interpolated PP produce very similar charge densities for β-Sn. It should be mentioned that the larger charge density changes could be a result of the mismatch between lattice constants obtained with the SLPP and the one computed by our local PP. Since the calculations use the same exchange-correlation functional and assuming that the SLPP is about as accurate as all-electron calculations, and since the optimization strives to match the experimental lattice constant (Eq. 15), the resulting electronic density belongs to a slightly compressed solid even if the experimental lattice constant is matched by the GPR-optimized PP compared to the SLPPs' results. Indeed, the charge density difference between GTH PP and LPPv2 PP reduces from 5% to 3.8% for α-Sn if the LPPv2 PP determined lattice constant is imposed on the GTH PP charge density calculation.

**Single atom properties**

It is also interesting to note that the properties of the isolated atoms reflect the insights from bulk calculations. The single atom density differences are also given in Table 2. Differences among SLPPs are small, less than 2%. This also applies to those among LPPs. In contrast, the differences are much larger (2-3%) when comparing LPPs and SLPPs. However, the density is surprisingly best reproduced by the interpolated LPP, which could be a consequence of the optimization scheme that averages over the property reproducibility from bulk and from single atom calculations. It hence increases accuracy for bulk materials while allowing a reduced accuracy for a single atom.

The eigenvalue spectrum gives information on the electronic structure. There are four electrons in a single Sn pseudoatom that occupy four states, of which three are degenerate. Due to the computational occupation scheme, two electrons are equally occupying the latter three degenerated states and two the other one, which is deeper in energy for all tested potentials, corresponding to 5p and 5s states in Sn, respectively. Hence the energy of the triple-degenerated states and the one directly above them define here the energy gap. ONV computed an energy gap between these levels of 3.2 eV, SLPPs of 3.0 eV and the LPPs 3.2 to 3.1 eV where the interpolated LPP obtained the largest and the LPPv1 the smallest energy gap. These values match reasonably well. This agrees with the insights from the bulk property analysis; that is: LPPv1 reproduces electronic properties the best (measured on SLPP results) among the here tested LPPs. Also, a contraction of electronic states was observed. The contraction increases with distance to the highest occupied molecular orbital. This contraction reaches ca. 1.1 eV for the occupied states corresponding to the 5$s$ orbital in a single Sn atom all LPPs.

Overall, generating an accurate local pseudopotential is a complex task for which a simple representation of a one-dimensional function is a severe limitation that is expressed through the, in principle, required knowledge of a general functional form, which is here avoided by using GPR, and furthermore complicated through at least three degrees of freedom during PP construction/optimization: that are the separation radius between core and valence region, the number of electrons included as valence states (which should be minimized) and the physical quantities for which the potential is being optimized.

Table 2. Relative density differences $\delta\rho$ between densities obtained with different semi-local and the interpolated local PP for α- and β-Sn computed with σ=5 meV and a dense k-point mesh (k × $a_0$ of ca. 80 Å).

|  |  | FHI | GTH | Interpolated local PP | GPR PP (LPPv1) | GPR PP (LPPv2) |
|---|---|---|---|---|---|---|
| α-Sn | FHI | N.A. | 1.92 % | 3.37 % | 2.85 % | 3.86 % |
|  | GTH | 1.92 % | N.A. | 4.60 % | 4.14 % | 5.06 % |
|  | Interpolated local PP[a] | 3.37 % | 4.60 % | N.A. | 0.82 % | 0.74 % |
|  | LPPv1 | 2.85 % | 4.14 % | 0.82 % | N.A. | 1.11 % |
|  | LPPv2 | 3.86 % | 5.06 % | 0.74 % | 1.11 % | N.A. |
| β-Sn | FHI | N.A. | 2.78 % | 1.72 % | 4.28 % | 2.80 % |
|  | GTH | 2.78 % | N.A. | 4.16 % | 3.02 % | 2.55 % |
|  | Interpolated local PP[a] | 1.72 % | 4.16 % | N.A. | 5.66 % | 4.01 % |
|  | LPPv1 | 4.28 % | 3.02 % | 5.66 % | N.A. | 1.67 % |
|  | LPPv2 | 2.80 % | 2.55 % | 4.01 % | 1.67 % | N.A. |
| Single atom | FHI | N.A. | 1.56 % | 2.05 % | 2.52 % | 2.50 % |
|  | GTH | 1.56 % | N.A. | 2.45 % | 2.86 % | 2.88 % |
|  | Interpolated local PP[a] | 2.05 % | 2.45 % | N.A. | 0.63 % | 0.56 % |
|  | LPPv1 | 2.52 % | 2.86 % | 0.63 % | N.A. | 0.18 % |
|  | LPPv2 | 2.50 % | 2.88 % | 0.56 % | 0.18 % | N.A. |

[a] Taken as an interpolation from local pseudopotential of In and Sb.[73,88]

Besides showing that structural and electronic properties can be simultaneously used to optimize a local PP and employing new representation math (GPR), we demonstrate that the core-valence separation radius can have a significant impact on electronic (bandgap), charge density and energetic properties (relative cohesive energy) while structural features are less affected. Especially, the charge density is a fundamental part of future OF-DFT developments and great caution needs to be taken in the pursue of accurate local PP and more accurate KEFs as the here presented results show that charge density differences can be as twice as large compared to differences seen among semi-local pseudopotentials.

## Conclusion

We present a new representation of local pseudopotentials, that is Gaussian Processes, i.e. a regression method known in the framework of machine learning, on the example of tin (Sn). With this new representation, we yield bulk material properties of the accuracy and precision previously only reported with non/semi-local pseudopotentials. Gaussian Processes has the advantage of being independent of a functional form including empirical parameters and it provides in combination with variational methods a direct way to optimize local pseudopotentials beyond most restrictions from DFT. These points are essential for i) further development of local-pseudopotential and ii) for the developing Kinetic Energy Functionals used in Orbital-Free Density-Functional Theory.

Errors in observables, including charge density, computed with local pseudopotentials can be related to two main effects: (i) the inadequacy of the LPP approximation itself, (ii) the non-optimality of the LPP shape. This work helps address (ii); while the LPPs produced here are likely improvable, with GPR, it is *in principle* possible to obtain the best possible LPP shape. We therefore hope that the use of GPR in particular and of non-parametric machine learning techniques in general will become a fruitful avenue for the production of local pseudopotentials for Orbital-free DFT.

## Acknowledgment

J.L. appreciates the financial support by the Ministry of Science and Technology (MOST) Taiwan under grant MOST108-2112-M-110-001-MY2 and acknowledges the National Center for High-performance Computing for computer time and facilities.

S. M. is grateful to Prof. Emily A. Carter for discussions and interactions with Carter groups which inspired us to apply machine learning to the problem of constructing local pseudopotentials suitable for orbital-free DFT.

# References

(1) Hohenberg, P.; Kohn, W. Inhomogeneous Electron Gas. *Phys. Rev.* **1964**, *136* (3B), B864–B871. https://doi.org/10.1103/PhysRev.136.B864.

(2) Fermi, E. Un Metodo Statistico per La Determinazione Di Alcune Prioprietà Dell'Atomo. *Rend. Accad. Naz. Lincei.* **1927**, *6*, 602.

(3) Thomas, L. H. The Calculation of Atomic Fields. *Math. Proc. Cambridge Philos. Soc.* **1927**, *23* (05), 542. https://doi.org/10.1017/S0305004100011683.

(4) Kohn, W.; Sham, L. J. Self-Consistent Equations Including Exchange and Correlation Effects. *Phys. Rev.* **1965**, *140* (4A), A1133–A1138. https://doi.org/10.1103/PhysRev.140.A1133.

(5) Becke, A. D. A New Mixing of Hartree–Fock and Local Density-functional Theories. *J. Chem. Phys.* **1993**, *98* (2), 1372–1377. https://doi.org/10.1063/1.464304.

(6) Kresse, G.; Joubert, D. From Ultrasoft Pseudopotentials to the Projector Augmented-Wave Method. *Phys. Rev. B* **1999**, *59* (3), 1758–1775. https://doi.org/10.1103/PhysRevB.59.1758.

(7) Blöchl, P. E. Projector Augmented-Wave Method. *Phys. Rev. B* **1994**, *50* (24), 17953–17979. https://doi.org/10.1103/PhysRevB.50.17953.

(8) Phillips, J. C.; Kleinman, L. New Method for Calculating Wave Functions in Crystals and Molecules. *Phys. Rev.* **1959**, *116* (2), 287–294. https://doi.org/10.1103/PhysRev.116.287.

(9) Gonze, X.; Stumpf, R.; Scheffler, M. Analysis of Separable Potentials. *Phys. Rev. B* **1991**, *44* (16), 8503–8513. https://doi.org/10.1103/PhysRevB.44.8503.

(10) Lejaeghere, K.; Bihlmayer, G.; Bjorkman, T.; Blaha, P.; Blugel, S.; Blum, V.; Caliste, D.; Castelli, I. E.; Clark, S. J.; Dal Corso, A.; de Gironcoli, S.; Deutsch, T.; Dewhurst, J. K.; Di Marco, I.; Draxl, C.; Du ak, M.; Eriksson, O.; Flores-Livas, J. A.; Garrity, K. F.; Genovese, L.; Giannozzi, P.; Giantomassi, M.; Goedecker, S.; Gonze, X.; Granas, O.; Gross, E. K. U.; Gulans, A.; Gygi, F.; Hamann, D. R.; Hasnip, P. J.; Holzwarth, N. A. W.; Iu an, D.; Jochym, D. B.; Jollet, F.; Jones, D.; Kresse, G.; Koepernik, K.; Kucukbenli, E.; Kvashnin, Y. O.; Locht, I. L. M.; Lubeck, S.; Marsman, M.; Marzari, N.; Nitzsche, U.; Nordstrom, L.; Ozaki, T.; Paulatto, L.; Pickard, C. J.; Poelmans, W.; Probert, M. I. J.; Refson, K.; Richter, M.; Rignanese, G.-M.; Saha, S.; Scheffler, M.; Schlipf, M.; Schwarz, K.; Sharma, S.; Tavazza, F.; Thunstrom, P.; Tkatchenko, A.; Torrent, M.; Vanderbilt, D.; van Setten, M. J.; Van Speybroeck, V.; Wills, J. M.; Yates, J. R.; Zhang, G.-X.; Cottenier, S.

Reproducibility in Density Functional Theory Calculations of Solids. *Science*. **2016**, *351* (6280), aad3000--aad3000. https://doi.org/10.1126/science.aad3000.

(11) Soler, J. M.; Artacho, E.; Gale, J. D.; García, A.; Junquera, J.; Ordejón, P.; Sánchez-Portal, D. The SIESTA Method for Ab Initio Order- N Materials Simulation. *J. Phys. Condens. Matter* **2002**, *14* (11), 2745–2779. https://doi.org/10.1088/0953-8984/14/11/302.

(12) Aarons, J.; Sarwar, M.; Thompsett, D.; Skylaris, C.-K. Perspective: Methods for Large-Scale Density Functional Calculations on Metallic Systems. *J. Chem. Phys.* **2016**, *145* (22), 220901. https://doi.org/10.1063/1.4972007.

(13) Mohr, S.; Eixarch, M.; Amsler, M.; Mantsinen, M. J.; Genovese, L. Linear Scaling DFT Calculations for Large Tungsten Systems Using an Optimized Local Basis. *Nucl. Mater. Energy* **2018**, *15*, 64–70. https://doi.org/10.1016/j.nme.2018.01.002.

(14) Goedecker, S. Electronic Structure Methods Exhibiting Linear Scaling of the Computational Effort with Respect to the Size of the System. **1998**, 1–73.

(15) Mao, Y.; Horn, P. R.; Mardirossian, N.; Head-Gordon, T.; Skylaris, C.-K.; Head-Gordon, M. Approaching the Basis Set Limit for DFT Calculations Using an Environment-Adapted Minimal Basis with Perturbation Theory: Formulation, Proof of Concept, and a Pilot Implementation. *J. Chem. Phys.* **2016**, *145* (4), 44109. https://doi.org/10.1063/1.4959125.

(16) Fonseca Guerra, C.; Snijders, J. G.; te Velde, G.; Baerends, E. J. Towards an Order- N DFT Method. *Theor. Chem. Accounts Theory, Comput. Model. (Theoretica Chim. Acta)* **1998**, *99* (6), 391–403. https://doi.org/10.1007/s002140050353.

(17) Hine, N. D. M.; Haynes, P. D.; Mostofi, A. A.; Skylaris, C.-K.; Payne, M. C. Linear-Scaling Density-Functional Theory with Tens of Thousands of Atoms: Expanding the Scope and Scale of Calculations with ONETEP. *Comput. Phys. Commun.* **2009**, *180* (7), 1041–1053. https://doi.org/10.1016/j.cpc.2008.12.023.

(18) Scuseria, G. E. Linear Scaling Density Functional Calculations with Gaussian Orbitals. *J. Phys. Chem. A* **1999**, *103* (25), 4782–4790. https://doi.org/10.1021/jp990629s.

(19) Hernández, E.; Gillan, M. J. Self-Consistent First-Principles Technique with Linear Scaling. *Phys. Rev. B* **1995**, *51* (15), 10157–10160. https://doi.org/10.1103/PhysRevB.51.10157.

(20) Hernández, E.; Gillan, M. J.; Goringe, C. M. Linear-Scaling Density-Functional-Theory

Technique: The Density-Matrix Approach. *Phys. Rev. B* **1996**, *53* (11), 7147–7157. https://doi.org/10.1103/PhysRevB.53.7147.

(21) *Computational Methods in Catalysis and Materials Science*; van Santen, R. A., Sautet, P., Eds.; Wiley-VCH Verlag GmbH & Co. KGaA: Weinheim, Germany, 2009. https://doi.org/10.1002/9783527625482.

(22) Witt, W. C.; del Rio, B. G.; Dieterich, J. M.; Carter, E. A. Orbital-Free Density Functional Theory for Materials Research. *J. Mater. Res.* **2018**, *33* (7), 777–795. https://doi.org/10.1557/jmr.2017.462.

(23) Shin, I.; Carter, E. A. Enhanced von Weizsäcker Wang-Govind-Carter Kinetic Energy Density Functional for Semiconductors. *J. Chem. Phys.* **2014**, *140* (18). https://doi.org/10.1063/1.4869867.

(24) Xia, J.; Carter, E. A. Single-Point Kinetic Energy Density Functionals: A Pointwise Kinetic Energy Density Analysis and Numerical Convergence Investigation. *Phys. Rev. B* **2015**, *91* (4), 045124. https://doi.org/10.1103/PhysRevB.91.045124.

(25) Karasiev, V. V; Jones, R. S.; Trickey, S. B.; Harris, F. E. Properties of Constraint-Based Single-Point Approximate Kinetic Energy Functionals. *Phys. Rev. B - Condens. Matter Mater. Phys.* **2009**, *80* (24), 1–17. https://doi.org/10.1103/PhysRevB.80.245120.

(26) Sim, E.; Larkin, J.; Burke, K.; Bock, C. W. Testing the Kinetic Energy Functional: Kinetic Energy Density as a Density Functional. *J. Chem. Phys.* **2003**, *118* (18), 8140–8148. https://doi.org/10.1063/1.1565316.

(27) Ke, Y.; Libisch, F.; Xia, J.; Wang, L. W.; Carter, E. A. Angular-Momentum-Dependent Orbital-Free Density Functional Theory. *Phys. Rev. Lett.* **2013**, *111* (6), 1–5. https://doi.org/10.1103/PhysRevLett.111.066402.

(28) Ludeña, E. V; Karasiev, V. V. KINETIC ENERGY FUNCTIONALS: HISTORY, CHALLENGES AND PROSPECTS. In *Reviews of Modern Quantum Chemistry*; WORLD SCIENTIFIC, 2002; pp 612–665. https://doi.org/10.1142/9789812775702_0022.

(29) Espinosa Leal, L. A.; Karpenko, A.; Caro, M. A.; Lopez-Acevedo, O. Optimizing a Parametrized Thomas–Fermi–Dirac–Weizsäcker Density Functional for Atoms. *Phys. Chem. Chem. Phys.* **2015**, *17* (47), 31463–31471. https://doi.org/10.1039/C5CP01211B.

(30) Constantin, L.; Fabiano, E.; Della Sala, F. Kinetic and Exchange Energy Densities near the

Nucleus. *Computation* **2016**, *4* (2), 19. https://doi.org/10.3390/computation4020019.

(31) Finzel, K. Shell-Structure-Based Functionals for the Kinetic Energy. *Theor. Chem. Acc.* **2015**, *134* (9), 1–10. https://doi.org/10.1007/s00214-015-1711-x.

(32) Karasiev, V. V.; Trickey, S. B. Issues and Challenges in Orbital-Free Density Functional Calculations. *Comput. Phys. Commun.* **2012**, *183* (12), 2519–2527. https://doi.org/10.1016/j.cpc.2012.06.016.

(33) Manzhos, S.; Golub, P. Data-Driven Kinetic Energy Density Fitting for Orbital-Free DFT : Linear vs Gaussian Process Regression.

(34) Golub, P.; Manzhos, S. Kinetic Energy Densities Based on the Fourth Order Gradient Expansion: Performance in Different Classes of Materials and Improvement via Machine Learning. *Phys. Chem. Chem. Phys.* **2019**, *21* (1), 378–395. https://doi.org/10.1039/C8CP06433D.

(35) Legrain, F.; Manzhos, S. Highly Accurate Local Pseudopotentials of Li, Na, and Mg for Orbital Free Density Functional Theory. *Chem. Phys. Lett.* **2015**, *622*, 99–103. https://doi.org/10.1016/j.cplett.2015.01.016.

(36) Huang, C.; Carter, E. A. Transferable Local Pseudopotentials for Magnesium, Aluminum and Silicon. *Phys. Chem. Chem. Phys.* **2008**, *10* (47), 7109–7120. https://doi.org/10.1039/b810407g.

(37) Zhou, B.; Carter, E. A. First Principles Local Pseudopotential for Silver: Towards Orbital-Free Density-Functional Theory for Transition Metals. *J. Chem. Phys.* **2005**, *122* (18), 184108. https://doi.org/10.1063/1.1897379.

(38) Pollack, L.; Perdew, J.; He, J.; Marques, M.; Nogueira, F.; Fiolhais, C. Tests of a Density-Based Local Pseudopotential for Sixteen Simple Metals. *Phys. Rev. B - Condens. Matter Mater. Phys.* **1997**, *55* (23), 15544–15551. https://doi.org/10.1103/PhysRevB.55.15544.

(39) Del Rio, B. G.; Dieterich, J. M.; Carter, E. A. Globally-Optimized Local Pseudopotentials for (Orbital-Free) Density Functional Theory Simulations of Liquids and Solids. *J. Chem. Theory Comput.* **2017**, *13* (8), 3684–3695. https://doi.org/10.1021/acs.jctc.7b00565.

(40) Fiolhais, C.; Perdew, J. P.; Armster, S. Q.; MacLaren, J. M.; Brajczewska, M. Dominant Density Parameters and Local Pseudopotentials for Simple Metals. *Phys. Rev. B* **1995**, *51* (20), 14001–14011. https://doi.org/10.1103/PhysRevB.51.14001.

(41) Mi, W.; Zhang, S.; Wang, Y.; Ma, Y.; Miao, M. First-Principle Optimal Local Pseudopotentials

Construction via Optimized Effective Potential Method. *J. Chem. Phys.* **2016**, *144* (13), 134108. https://doi.org/10.1063/1.4944989.

(42) Hoshino, K.; Young, W. H. A Simple Local Pseudopotential for Lithium. *J. Phys. F Met. Phys.* **1986**, *16* (11), 1659–1670. https://doi.org/10.1088/0305-4608/16/11/007.

(43) Jones, D. E.; Ghandehari, H.; Facelli, J. C. A Review of the Applications of Data Mining and Machine Learning for the Prediction of Biomedical Properties of Nanoparticles. *Comput. Methods Programs Biomed.* **2016**, *132*, 93–103. https://doi.org/10.1016/j.cmpb.2016.04.025.

(44) Liu, Y.; Zhao, T.; Ju, W.; Shi, S. Materials Discovery and Design Using Machine Learning. *J. Mater.* **2017**, *3* (3), 159–177. https://doi.org/10.1016/j.jmat.2017.08.002.

(45) Kamath, A.; Vargas-Hernández, R. A.; Krems, R. V; Carrington, T.; Manzhos, S. Neural Networks vs Gaussian Process Regression for Representing Potential Energy Surfaces: A Comparative Study of Fit Quality and Vibrational Spectrum Accuracy. *J. Chem. Phys.* **2018**, *148* (24), 241702. https://doi.org/10.1063/1.5003074.

(46) Kolb, B.; Marshall, P.; Zhao, B.; Jiang, B.; Guo, H. Representing Global Reactive Potential Energy Surfaces Using Gaussian Processes. *J. Phys. Chem. A* **2017**, *121* (13), 2552–2557. https://doi.org/10.1021/acs.jpca.7b01182.

(47) Behler, J. Perspective: Machine Learning Potentials for Atomistic Simulations. *J. Chem. Phys.* **2016**, *145* (17), 170901. https://doi.org/10.1063/1.4966192.

(48) Rupp, M. Machine Learning for Quantum Mechanics in a Nutshell. *Int. J. Quantum Chem.* **2015**, *115* (16), 1058–1073. https://doi.org/10.1002/qua.24954.

(49) Osborne, M. A.; Garnett, R.; Roberts, S. J. Gaussian Processes for Global Optimization. *3rd Int. Conf. Learn. Intell. Optim. LION3* **2009**, No. x, 1–15.

(50) Li, L.; Snyder, J. C.; Pelaschier, I. M.; Huang, J.; Niranjan, U. N.; Duncan, P.; Rupp, M.; Müller, K. R.; Burke, K. Understanding Machine-Learned Density Functionals. *Int. J. Quantum Chem.* **2016**, *116* (11), 819–833. https://doi.org/10.1002/qua.25040.

(51) Manzhos, S. Machine Learning for the Solution of the Schrödinger Equation. *Mach. Learn. Sci. Technol.* **2020**, *1* (1), 013002. https://doi.org/10.1088/2632-2153/ab7d30.

(52) Brockherde, F.; Vogt, L.; Li, L.; Tuckerman, M. E.; Burke, K.; Müller, K. R. Bypassing the Kohn-Sham Equations with Machine Learning. *Nat. Commun.* **2017**, *8* (1).

https://doi.org/10.1038/s41467-017-00839-3.

(53) Fujinami, M.; Kageyama, R.; Seino, J.; Ikabata, Y.; Nakai, H. Orbital-Free Density Functional Theory Calculation Applying Semi-Local Machine-Learned Kinetic Energy Density Functional and Kinetic Potential. *Chem. Phys. Lett.* **2020**, *748*, 137358. https://doi.org/10.1016/j.cplett.2020.137358.

(54) Seino, J.; Kageyama, R.; Fujinami, M.; Ikabata, Y.; Nakai, H. Semi-Local Machine-Learned Kinetic Energy Density Functional Demonstrating Smooth Potential Energy Curves. *Chem. Phys. Lett.* **2019**, *734* (August), 136732. https://doi.org/10.1016/j.cplett.2019.136732.

(55) Rasmussen, C. E.; Williams, C. K. I.; Processes, G.; Press, M. I. T.; Jordan, M. I.; Carl Edward Rasmussen, C. K. I. W. *Gaussian Processes for Machine Learning*; MIT Press: Cambridge MA, 2006; Vol. 14. https://doi.org/10.1142/S0129065704001899.

(56) Sacks, J.; Schiller, S. B.; Welch, W. J. Designs for Computer Experiments. *Technometrics* **1989**, *31* (1), 41–47. https://doi.org/10.1080/00401706.1989.10488474.

(57) Legrain, F.; Manzhos, S. Understanding the Difference in Cohesive Energies between Alpha and Beta Tin in DFT Calculations. *AIP Adv.* **2016**, *6* (4), 45116. https://doi.org/10.1063/1.4948434.

(58) Legrain, F.; Malyi, O. I.; Manzhos, S. A Comparative Computational Study of Li, Na, and Mg Insertion in $α$-Sn. *MRS Proc.* **2014**, *1678*. https://doi.org/10.1557/opl.2014.743.

(59) Oehl, N.; Hardenberg, L.; Knipper, M.; Kolny-Olesiak, J.; Parisi, J.; Plaggenborg, T. Critical Size for the β- to α-Transformation in Tin Nanoparticles after Lithium Insertion and Extraction. *CrystEngComm* **2015**, *17* (19), 3695–3700. https://doi.org/10.1039/C5CE00148J.

(60) Bauer, M.; Taraci, J.; Tolle, J.; Chizmeshya, A. V. G.; Zollner, S.; Smith, D. J.; Menendez, J.; Hu, C.; Kouvetakis, J. Ge–Sn Semiconductors for Band-Gap and Lattice Engineering. *Appl. Phys. Lett.* **2002**, *81* (16), 2992–2994. https://doi.org/10.1063/1.1515133.

(61) Tan, C. G.; Zhou, P.; Lin, J. G.; Sun, L. Z. Two-Dimensional Semiconductors XY 2 (X = Ge,Sn;Y = S,Se) with Promising Piezoelectric Properties. *Comput. Condens. Matter* **2017**, *11*, 33–39. https://doi.org/10.1016/j.cocom.2017.04.001.

(62) Taraci, J.; Zollner, S.; McCartney, M. R.; Menendez, J.; Santana-Aranda, M. A.; Smith, D. J.; Haaland, A.; Tutukin, A. V; Gundersen, G.; Wolf, G.; Kouvetakis, J. Synthesis of Silicon-Based Infrared Semiconductors in the Ge−Sn System Using Molecular Chemistry Methods. *J. Am. Chem.*

*Soc.* **2001**, *123* (44), 10980–10987. https://doi.org/10.1021/ja0115058.

(63) Kunzler, J. E.; Buehler, E.; Hsu, F. S. L.; Wernick, J. H. Superconductivity in Nb3 Sn at High Current Density in a Magnetic Field of 88 Kgauss. *Phys. Rev. Lett.* **1961**, *6* (3), 89–91. https://doi.org/10.1103/PhysRevLett.6.89.

(64) Watanabe, K.; Yamada, Y.; Sakuraba, J.; Hata, F.; Chong, C. K.; Hasebe, T.; Ishihara, M. (Nb, Ti) 3 Sn Superconducting Magnet Operated at 11 K in Vacuum Using High- T c (Bi, Pb) 2 Sr 2 Ca 2 Cu 3 O 10 Current Leads. *Jpn. J. Appl. Phys.* **1993**, *32* (Part 2, No. 4A), L488–L490. https://doi.org/10.1143/JJAP.32.L488.

(65) Lemaignan, C.; Motta, A. T. Zirconium Alloys in Nuclear Applications. In *Materials Science and Technology*; Wiley-VCH Verlag GmbH & Co. KGaA: Weinheim, Germany, 2006. https://doi.org/10.1002/9783527603978.mst0111.

(66) Olander, D. Nuclear Fuels – Present and Future. *J. Nucl. Mater.* **2009**, *389* (1), 1–22. https://doi.org/10.1016/j.jnucmat.2009.01.297.

(67) Sierros, K. A.; Morris, N. J.; Ramji, K.; Cairns, D. R. Stress–Corrosion Cracking of Indium Tin Oxide Coated Polyethylene Terephthalate for Flexible Optoelectronic Devices. *Thin Solid Films* **2009**, *517* (8), 2590–2595. https://doi.org/10.1016/j.tsf.2008.10.031.

(68) Rakspun, J.; Kantip, N.; Vailikhit, V.; Choopun, S.; Tubtimtae, A. Multi-Phase Structures of Boron-Doped Copper Tin Sulfide Nanoparticles Synthesized by Chemical Bath Deposition for Optoelectronic Devices. *J. Phys. Chem. Solids* **2018**, *115*, 103–112. https://doi.org/10.1016/j.jpcs.2017.12.028.

(69) Mohammed, D. W.; Ameen, R. B.; Sierros, K. A.; Bowen, J.; Kukureka, S. N. Twisting Fatigue in Multilayer Films of Ag-Alloy with Indium Tin Oxide on Polyethylene Terephthalate for Flexible Electronics Devices. *Thin Solid Films* **2018**, *645*, 241–252. https://doi.org/10.1016/j.tsf.2017.10.047.

(70) Houben, K.; Jochum, J. K.; Lozano, D. P.; Bisht, M.; Menéndez, E.; Merkel, D. G.; Rüffer, R.; Chumakov, A. I.; Roelants, S.; Partoens, B.; Milošević, M. V.; Peeters, F. M.; Couet, S.; Vantomme, A.; Temst, K.; Van Bael, M. J. In Situ Study of the $\alpha$-Sn to $\beta$-Sn Phase Transition in Low-Dimensional Systems: Phonon Behavior and Thermodynamic Properties. *Phys. Rev. B* **2019**, *100* (7), 075408. https://doi.org/10.1103/PhysRevB.100.075408.

(71) Haq, A. U.; Askari, S.; McLister, A.; Rawlinson, S.; Davis, J.; Chakrabarti, S.; Svrcek, V.; Maguire,

P.; Papakonstantinou, P.; Mariotti, D. Size-Dependent Stability of Ultra-Small α-/β-Phase Tin Nanocrystals Synthesized by Microplasma. *Nat. Commun.* **2019**, *10* (1), 817. https://doi.org/10.1038/s41467-019-08661-9.

(72) Neal, R. M. Priors for Infinite Networks. In *Bayesian Learning for Neural Networks*; Springer New York: New York, NY, 1996; pp 29–53. https://doi.org/10.1007/978-1-4612-0745-0_2.

(73) Huang, C.; Carter, E. A. Nonlocal Orbital-Free Kinetic Energy Density Functional for Semiconductors. *Phys. Rev. B - Condens. Matter Mater. Phys.* **2010**, *81* (4), 1–15. https://doi.org/10.1103/PhysRevB.81.045206.

(74) Xia, J.; Huang, C.; Shin, I.; Carter, E. A. Can Orbital-Free Density Functional Theory Simulate Molecules. *J. Chem. Phys.* **2012**, *136* (8). https://doi.org/10.1063/1.3685604.

(75) Pedregosa, F.; Varoquaux, G.; Gramfort, A.; Michel, V.; Thirion, B.; Grisel, O.; Blondel, M.; Louppe, G.; Prettenhofer, P.; Weiss, R.; Dubourg, V.; Vanderplas, J.; Passos, A.; Cournapeau, D.; Brucher, M.; Perrot, M.; Duchesnay, É. Scikit-Learn: Machine Learning in Python. *J. Mach. Learn. Res.* **2012**, *12*, 2825–2830.

(76) Gonze, X.; Jollet, F.; Abreu Araujo, F.; Adams, D.; Amadon, B.; Applencourt, T.; Audouze, C.; Beuken, J.-M.; Bieder, J.; Bokhanchuk, A.; Bousquet, E.; Bruneval, F.; Caliste, D.; Côté, M.; Dahm, F.; Da Pieve, F.; Delaveau, M.; Di Gennaro, M.; Dorado, B.; Espejo, C.; Geneste, G.; Genovese, L.; Gerossier, A.; Giantomassi, M.; Gillet, Y.; Hamann, D. R.; He, L.; Jomard, G.; Laflamme Janssen, J.; Le Roux, S.; Levitt, A.; Lherbier, A.; Liu, F.; Lukačević, I.; Martin, A.; Martins, C.; Oliveira, M. J. T.; Poncé, S.; Pouillon, Y.; Rangel, T.; Rignanese, G.-M.; Romero, A. H.; Rousseau, B.; Rubel, O.; Shukri, A. A.; Stankovski, M.; Torrent, M.; Van Setten, M. J.; Van Troeye, B.; Verstraete, M. J.; Waroquiers, D.; Wiktor, J.; Xu, B.; Zhou, A.; Zwanziger, J. W. Recent Developments in the ABINIT Software Package. *Comput. Phys. Commun.* **2016**, *205*, 106–131. https://doi.org/10.1016/j.cpc.2016.04.003.

(77) Gonze, X.; Amadon, B.; Anglade, P.-M.; Beuken, J.-M.; Bottin, F.; Boulanger, P.; Bruneval, F.; Caliste, D.; Caracas, R.; Côté, M.; Deutsch, T.; Genovese, L.; Ghosez, P.; Giantomassi, M.; Goedecker, S.; Hamann, D. R.; Hermet, P.; Jollet, F.; Jomard, G.; Leroux, S.; Mancini, M.; Mazevet, S.; Oliveira, M. J. T.; Onida, G.; Pouillon, Y.; Rangel, T.; Rignanese, G.-M.; Sangalli, D.; Shaltaf, R.; Torrent, M.; Verstraete, M. J.; Zerah, G.; Zwanziger, J. W. ABINIT: First-Principles Approach to Material and Nanosystem Properties. *Comput. Phys. Commun.* **2009**, *180* (12), 2582–2615. https://doi.org/10.1016/j.cpc.2009.07.007.

(78) Gonze, X. A Brief Introduction to the ABINIT Software Package. *Zeitschrift für Krist. - Cryst. Mater.* **2005**, *220* (5/6). https://doi.org/10.1524/zkri.220.5.558.65066.

(79) Bottin, F.; Leroux, S.; Knyazev, A.; Zérah, G. Large-Scale Ab Initio Calculations Based on Three Levels of Parallelization. *Comput. Mater. Sci.* **2008**, *42* (2), 329–336. https://doi.org/10.1016/j.commatsci.2007.07.019.

(80) Hamann, D. R. Optimized Norm-Conserving Vanderbilt Pseudopotentials. *Phys. Rev. B* **2013**, *88* (8), 85117. https://doi.org/10.1103/PhysRevB.88.085117.

(81) Perdew, J. P.; Burke, K.; Ernzerhof, M. Generalized Gradient Approximation Made Simple. *Phys. Rev. Lett.* **1996**, *77* (18), 3865–3868. https://doi.org/10.1103/PhysRevLett.77.3865.

(82) Brownlee, L. D. Lattice Constant of Grey Tin. *Nature* **1950**, *166* (4220), 482–482. https://doi.org/10.1038/166482a0.

(83) Grey Tin (Alpha-Sn), Crystal Structure, Lattice Parameter (Pure Sn, Sn-IV Alloys), Thermal Expansion. In *Group IV Elements, IV-IV and III-V Compounds. Part b - Electronic, Transport, Optical and Other Properties*; Madelung, O., Rössler, U., Schulz, M., Eds.; Springer Berlin Heidelberg: Berlin, Heidelberg, 2002; pp 1–6. https://doi.org/10.1007/10832182_543.

(84) *Atomic Defects in Metals*; Ullmaier, H., Ed.; Landolt-Börnstein - Group III Condensed Matter; Springer-Verlag: Berlin/Heidelberg, 1991; Vol. 25. https://doi.org/10.1007/b37800.

(85) Jain, A.; Ong, S. P.; Hautier, G.; Chen, W.; Richards, W. D.; Dacek, S.; Cholia, S.; Gunter, D.; Skinner, D.; Ceder, G.; Persson, K. A. Commentary: The Materials Project: A Materials Genome Approach to Accelerating Materials Innovation. *APL Mater.* **2013**, *1* (1), 11002. https://doi.org/10.1063/1.4812323.

(86) Troullier, N.; Martins, J. L. Efficient Pseudopotentials for Plane-Wave Calculations. *Phys. Rev. B* **1991**, *43* (3), 1993–2006. https://doi.org/10.1103/PhysRevB.43.1993.

(87) Goedecker, S.; Teter, M.; Hutter, J. Separable Dual-Space Gaussian Pseudopotentials. *Phys. Rev. B* **1996**, *54* (3), 1703–1710. https://doi.org/10.1103/PhysRevB.54.1703.

(88) Chen, M.; Xia, J.; Huang, C.; Dieterich, J. M.; Hung, L.; Shin, I.; Carter, E. A. Introducing PROFESS 3.0: An Advanced Program for Orbital-Free Density Functional Theory Molecular Dynamics Simulations. *Comput. Phys. Commun.* **2015**, *190*, 228–230. https://doi.org/10.1016/j.cpc.2014.12.021.

(89) Rahm, M.; Hoffmann, R.; Ashcroft, N. W. Atomic and Ionic Radii of Elements 1–96. *Chem. - A Eur. J.* **2016**, *22* (41), 14625–14632. https://doi.org/10.1002/chem.201602949.

(90) Kresse, G.; Furthmüller, J. Efficiency of Ab-Initio Total Energy Calculations for Metals and Semiconductors Using a Plane-Wave Basis Set. *Comput. Mater. Sci.* **1996**, *6* (1), 15–50. https://doi.org/10.1016/0927-0256(96)00008-0.

(91) Kresse, G.; Hafner, J. Ab Initio Molecular Dynamics for Liquid Metals. *Phys. Rev. B* **1993**, *47* (1), 558–561. https://doi.org/10.1103/PhysRevB.47.558.

(92) Haas, P.; Tran, F.; Blaha, P. Calculation of the Lattice Constant of Solids with Semilocal Functionals. *Phys. Rev. B - Condens. Matter Mater. Phys.* **2009**, *79* (8), 1–10. https://doi.org/10.1103/PhysRevB.79.085104.

(93) Zaharioudakis, D. Tetrahedron Methods for Brillouin Zone Integration. *Comput. Phys. Commun.* **2004**, *157* (1), 17–31. https://doi.org/10.1016/S0010-4655(03)00489-2.

(94) Carrasco, R. A.; Zamarripa, C. M.; Zollner, S.; Menéndez, J.; Chastang, S. A.; Duan, J.; Grzybowski, G. J.; Claflin, B. B.; Kiefer, A. M. The Direct Bandgap of Gray α -Tin Investigated by Infrared Ellipsometry. *Appl. Phys. Lett.* **2018**, *113* (23), 232104. https://doi.org/10.1063/1.5053884.

(95) Ewald, A. W.; Kohnke, E. E. Measurements of Electrical Conductivity and Magnetoresistance of Gray Tin Filaments. *Phys. Rev.* **1955**, *97* (3), 607–613. https://doi.org/10.1103/PhysRev.97.607.